\documentclass[onecollarge,natbib]{svjour2}
\bibpunct{[}{]}{;}{n}{}{,} 
\smartqed  
\usepackage{graphicx, epstopdf, hyperref}
\usepackage[fleqn]{amsmath}
\journalname{Archive of Applied Mechanics}
\begin{document}

\title{Ratios of vector and pseudoscalar $B$  meson decay constants in the light-cone quark model \thanks{Authors would like to thank the Department of Science and Technology (Ref No. SB/S2/HEP-004/2013) Government of India for financial support.}
}


\author{Nisha Dhiman         \and
        Harleen Dahiya 
}


\institute{N. Dhiman \and H. Dahiya \at
              Dr. B. R. Ambedkar National Institute of Technology, Jalandhar, India \\
              \email{nishdhiman1292@gmail.com}    
}

\date{Received: date / Accepted: date}

\maketitle

\begin{abstract}
We study the decay constants of pseudoscalar and vector $B$ meson in the framework of light-cone quark model (LCQM). We apply the variational method to the relativistic Hamiltonian with the Gaussian-type trial wave function to obtain the values of $\beta$ (scale parameter). Then with the help of known values of constituent quark masses, we obtain the numerical results for the decay constants $f_P$ and $f_V$, respectively. We compare our numerical results with the existing experimental data.
\end{abstract}

\section{Introduction}
\label{intro}
The study of the decay constants of heavy mesons is very important, since it provides a direct source of information on the Cabibbo-Kobayashi-Maskawa (CKM) matrix elements which describe the couplings of the third generation of quarks to the lighter quarks. These matrix elements are the fundamental parameters of the Standard Model (SM) and their precise measurement will allow us to test the unitarity of the quark mixing matrix and $CP$ violation in the SM \cite{mona}. However, the uncertainty in the knowledge of the decay constant make it difficult to extract precisely the CKM matrix elements from the experimental data. For example, in the lowest order approximation, the decay widths of the pseudoscalar and vector mesons can be written as \cite{zhi}
\begin{eqnarray}
\label{eqn:1}
\Gamma (P \to \ell \nu) &=& \frac{G_F^2}{8 \pi}f_P^2 m_l^2 m_P \left(1 - \frac{m_{\ell}^2}{m_P^2}\right)^2 |V_{qQ}|^2, \quad {\rm and}\nonumber\\
\Gamma (V \to \ell \nu) &=& \frac{G_F^2}{12 \pi}f_V^2 m_V^3 \left(1 - \frac{m_{\ell}^2}{m_V^2}\right)^2 \left(1 + \frac{m_{\ell}^2}{2 m_V^2}\right) |V_{qQ}|^2.
\end{eqnarray}
The experimental measurements of purely leptonic decay branching fractions and their lifetimes allow us to determine the product $f_P |V_{qQ}|$ ($f_V |V_{qQ}|$). Thus, a precise theoretical input on $f_{P(V)}$ can allow a determination of the CKM matrix element. The theoretical calculations of the decay constants of $B$ mesons require non-perturbative treatment since at short distances, the interactions are dominated by strong force. There have been many theoretical groups that are working on the calculation of the decay constants in the realm of non-perturbative QCD using different models. Here we focus on one such method that is useful for solving non-perturbative problems of hadron physics: the light-cone quark model (LCQM) \cite{chien, chien1, chao, ho}. The LCQM has been widely used in the phenomenological study of meson physics. It provides an advantage of the equal light-cone time ($x^+ = x^0 +x^3$) quantization and includes the important relativistic effects that are neglected in the traditional constituent quark model \cite{brodsky1, lepage, dirac}. Moreover, vacuum fluctuations are absent in light-cone field theory and the state can be expanded in fock space in terms of frame independent n particle light-cone wave functions (LCWFs) \cite{brodsky2}. The LCWFs are independent of the hadron momentum and thus are explicitly Lorentz invariant \cite{brodsky3}. 
The present work is devoted to the analysis of the decay constants by fixing the scale parameter $\beta$ under the Martin potential \cite{martin}, Cornell potential \cite{cornell}, Logarithmic potential \cite{log} and the combination of harmonic and Yukawa potentials \cite{harmonic}, respectively within the light-cone framework.

\section{Decay constants for $B$ mesons in LCQM}
\label{sec:1}
The bound state of a heavy meson composed of a light quark $q$ and a heavy antiquark $\bar{Q}$ with total momentum $P$ and spin $S$ is represented as \cite{cai}
\begin{eqnarray}
\label{eqn:2}
|M(P,S,S_z)\rangle &=& \int\frac{dp_{q}^+d^2\textbf{p}_{q_{\bot}}}{16\pi^3} \frac{dp_{\bar Q}^+d^2\textbf{p}_{\bar Q_\bot}}{16\pi^3}16\pi^3 \delta^3(\tilde P-\tilde p_{q}-\tilde p_{\bar Q}) \nonumber\\ && \times \sum\limits_{\lambda_{q},\lambda_{\bar Q}}\Psi^{SS_z}(\tilde p_{q},\tilde p_{\bar Q},\lambda_{q},\lambda_{\bar Q}) \ 
 |q(p_{q},\lambda_{q})\bar Q(p_{\bar Q},\lambda_{\bar Q})\rangle,
\end{eqnarray}
where $p_{q (\bar Q)}$ and $\lambda_{q (\bar Q)}$ are the on-mass shell light-front momentum and the light-front helicity of the constituent quark (antiquark) respectively. The light-front momenta $p_{q}$ and $p_{\bar Q}$ in terms of light-cone variables are $p_{q}^+=x_1 P^+$, $p_{\bar Q}^+=x_2 P^+$, $\textbf{p}_{q_{\perp}}=x_1\textbf{P}_{\perp}+\textbf{k}_{\perp}$, $\textbf{p}_{\bar{Q}_\perp}=x_2\textbf{P}_{\perp}-\textbf{k}_{\perp}$,
where $x_{1 (2)}$ is the light-cone momentum fraction satisfying the relation $x_1 + x_2 = 1$ and $\textbf{k}_{\perp}$ is the relative transverse momentum of the constituent.
The momentum-space light-cone wave function $\Psi^{SS_z}$ in Eq. (\ref{eqn:2}) can be expressed as a covariant form \cite{chao, cai, ho1}
\begin{eqnarray}
\label{eqn:3}
        \Psi^{SS_z}(\tilde p_{q},\tilde p_{\bar Q},\lambda_{q},\lambda_{\bar Q})
                ={\sqrt{p_{q}^+p_{\bar Q}^+}\over \sqrt{2} ~{\sqrt{{M_0^2} - (m_{q} - m_{\bar Q})^2}}}
        ~\bar u(p_{q},\lambda_{q})\,\Gamma\, v(p_{\bar Q},\lambda_{\bar Q})~\sqrt{dk_z\over dx}~\phi(x, \textbf{k}_\bot),
\end{eqnarray}
where $\phi(x, \textbf{k}_\bot)$ describes the momentum distribution of the constituents in the bound state. We use the Gaussian-type wave function to describe the radial wave function $\phi$
\begin{eqnarray}
\label{eqn:4}
\phi(x,\textbf{k}_\bot)=\frac{1}{(\sqrt{\pi}\beta)^{3/2}}
\exp(-\textbf{k}^{2}/2\beta^{2}),
\end{eqnarray}
where $\beta$ represents the scale parameter and $\textbf{k}^2=\textbf{k}^2_\bot + k^2_z$ is the internal momentum of the meson.\\
The pseudoscalar and the vector meson decay constants are defined through the matrix elements of axial and vector currents between the meson state and the vacuum \cite{ho}, i.e.
\begin{eqnarray}
\label{eqn:5}
\langle 0|A^\mu|P(P) \rangle = if_P P^\mu, \ {\rm and} \ \  \langle 0|V^\mu|V (P) \rangle = f_V M_V \epsilon^{\mu}.
\end{eqnarray}
Using light-cone wave function, the decay constants of the pseudoscalar and vector mesons are given by
\begin{eqnarray}
\label{eqn:6}
f_P = 2\sqrt{6}\int\frac{dxd^2\textbf{k}_{\perp}}{2(2\pi)^3} \sqrt{dk_z\over dx}\phi(x,\textbf{k}_\perp) \frac{\mathcal{A}}{\sqrt{\mathcal{A}^2+\textbf{k}^2_{\perp}}},
\end{eqnarray}
\begin{eqnarray}
\label{eqn:7}
f_V &=& 2\sqrt{6}\int\frac{dxd^2\textbf{k}_\perp}{2(2\pi)^3} \sqrt{dk_z\over dx}\frac{\phi(x,\textbf{k}_\perp)}{\sqrt {{\cal A}^2+\textbf{k}_\perp^2}}\frac{1}{M_{0}}
\Big\{m_q m_{\bar Q}+x(1-x)M_{0}^2+\nonumber \\
&&\textbf{k}_\perp^2 +\frac{{\cal B}}{2W}\left[\frac{m_q^2+\textbf{k}_\perp^2}{1-x}-\frac{m_{\bar Q}^2+\textbf{k}_\perp^2}{x}-(1-2x)M_{0}^2\right]\Big\}.
\end{eqnarray}
where $\mathcal{A} = m_q x + m_{\bar Q} (1 - x)$, $\mathcal{B} = m_q x - m_{\bar Q} (1 - x)$ and $W = M_0 + m_q +m_{\bar Q}$. For a given value of $\beta$, $f_P$ and $f_V$ can be calculated from Eqs. (\ref{eqn:6}) and (\ref{eqn:7}) using the constituent quark masses of $u$, $d$, $s$ and $b$ quarks, respectively.

\section{Numerical results}
\label{sec:2}
\begin{figure}[h]
\centering
\minipage{0.95\linewidth}
  \includegraphics[width=2.8 in]{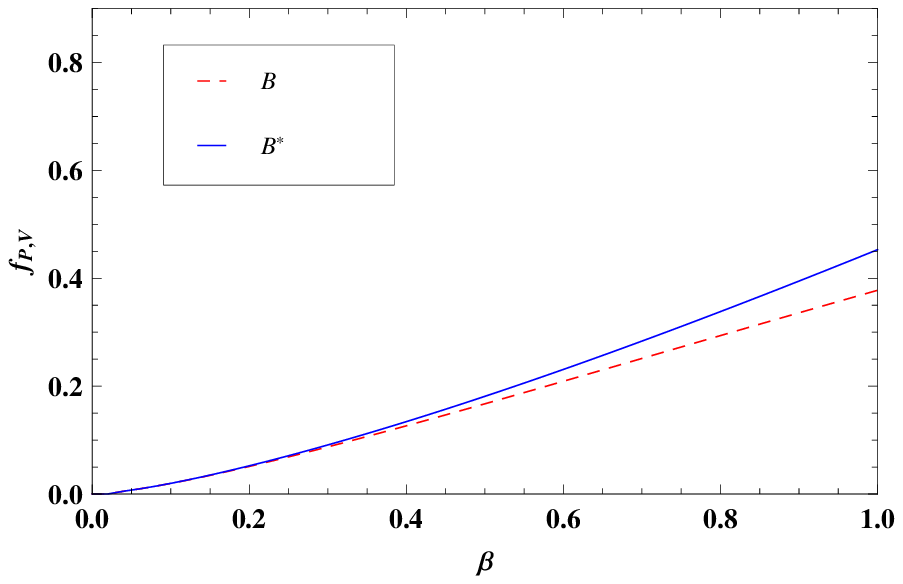}
  \includegraphics[width=2.8 in]{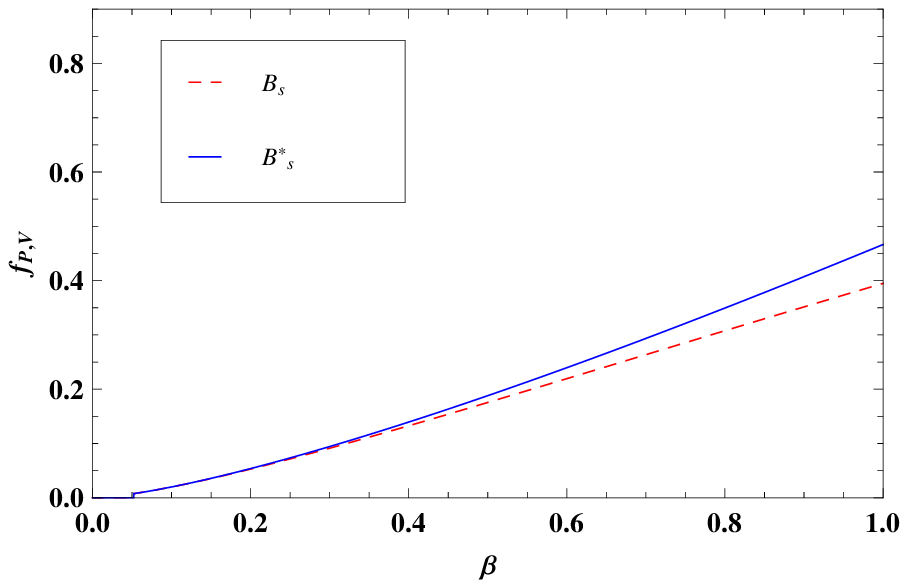}
\endminipage
\caption{$f_{B}$, $f_{B^*}$, $f_{B_s}$ and $f_{B^*_s}$ as functions of the parameter $\beta$ (in GeV).} 
\label{fig:1}
\end{figure}

\begin{figure}[h]
  \begin{center}
    \includegraphics[width=2.8 in]{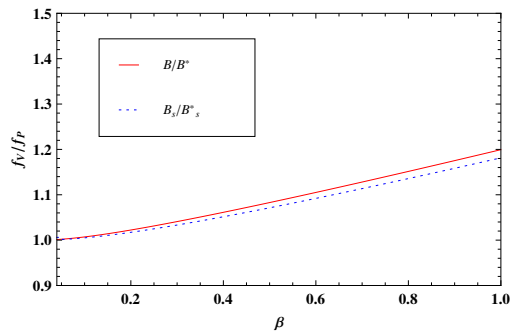}
  \end{center}
  \caption{$f_{B^*}/f_{B}$ and $f_{B^*_s}/f_{B_s}$ as functions of the parameter $\beta$ (in GeV).} 
  \label{fig:3}
\end{figure}

Numerically, we obtain the pseudoscalar and the vector meson decay constants for $B$ and $B_s$ mesons as functions of the scale parameter $\beta$, using the following values of constituent quark masses \cite{chao}
\begin{center}
$m_u = m_d = 0.25$ GeV, $m_s = 0.38$ GeV and $m_b = 4.8$ GeV.
\end{center}
We present the decay constants $f_P$ and $f_V$ as functions of $\beta$ for $B$ and $B_s$ mesons in Fig. \ref{fig:1}. Even though it is expected that the pseudoscalar and the vector meson decay constants will have similar values because they differ with each other only in their internal spin configurations, however, to exploit their quantitative difference, we present in Fig. \ref{fig:3} our results for the ratios $f_V/ f_P$ as a function of $\beta$.

We apply the variational method to the following relativistic Hamiltonian to obtain the precise values of $\beta$ \cite{hwang}:
\begin{eqnarray}
\label{eqn:8}
H = \sqrt{\textbf{k}^{2} + m^2_{q}} + \sqrt{\textbf{k}^{2} + m^2_{\bar Q}} + V (r).
\end{eqnarray}
We consider the Gaussian wave function in the above equation as our trial wave function with the variational parameter $\beta$:
\begin{eqnarray}
\label{eqn:9}
\phi(x,\textbf{k}_\bot)=\frac{1}{(\sqrt{\pi}\beta)^{3/2}}
\exp(-\textbf{k}^{2}/2\beta^{2}). 
\end{eqnarray}
We can obtain the ground state energy for our system by minimizing the expectation value of the Hamiltonian $H$, $\langle H\rangle = \langle\psi ({\bf r})|H|\psi ({\bf r})\rangle = E (\beta)$, that is,
\begin{eqnarray}
\label{eqn:10}
\frac{d E (\beta)}{d \beta} = 0 \quad {\rm at}\,\, \beta = \bar{\beta}, \nonumber
\end{eqnarray}
where $\bar{\beta}$ denotes the inverse size of the meson ($\langle r^2 \rangle^{1/2} = 3/(2 \bar{\beta})$) and $E(\bar \beta) = \bar E$ approximates the meson mass $m_M$.
Our results for the variational parameter $\beta$ of the Gaussian wave function in different potential models has been listed in Table \ref{tab1}.
\begin{table}[h]
\centering
\caption{Parameter $\beta$ and the corresponding values of minimum energy $E (\bar{\beta})$ (in units of GeV) in different potential models.}
\label{tab1}
\begin{tabular}{lclcl}
\hline\hline\\[-1.5 ex]
\multicolumn{1}{c}{Potential Model}                                                     & $\bar{\beta}_{B}$     & \multicolumn{1}{c}{$\bar{E}_{B}$} & $\bar{\beta}_{B_s}$     & \multicolumn{1}{c}{$\bar{E}_{B_s}$}
\\
\hline\\[-1.5 ex]
Martin \cite{martin}                                                                                   &\hspace{0.1in} 0.592 &\hspace{0.01in}    4.803                   &\hspace{0.01in}    0.628 &\hspace{0.01in}                     4.869  \\
Cornell \cite{cornell}                                                                                  &\hspace{0.1in} 0.561 &\hspace{0.01in}    5.624                      &\hspace{0.01in}    0.600 &\hspace{0.01in}    5.692  \\
Logarithmic \cite{log}  
&\hspace{0.1in} 0.595 &\hspace{0.01in}     5.311                     &\hspace{0.01in}    0.633 &\hspace{0.01in}      5.376   \\
Harmonic plus Yukawa \cite{harmonic} 
&\hspace{0.1in} 0.408 &\hspace{0.01in}    5.297                      &\hspace{0.01in}    0.496 &\hspace{0.01in}    5.436 \\
\hline\hline
\end{tabular}
\end{table}
Also, Table \ref{tab1} gives us the following average values of $\bar{E}$ which approximates the meson masses: $\bar{E}_{B} = 5.26$ GeV and $\bar{E}_{B_s} = 5.34$ GeV. \nonumber \\
The calculated values of $B$ meson masses are in reasonable agreement compared to the measured values ($m_B = 5.28$ GeV and $m_{B_s} = 5.37$ GeV) \cite{patri}.
The numerical results of the decay constants that we obtained and their ratios $f_{B^*}/f_{B}$, $f_{B_{s^*}}/f_{B_s}$ have been listed in Table \ref{tab2}.

\begin{table}[h]
\centering
\caption{Pseudoscalar and vector meson decay constants for $B$ mesons (in units of MeV) and their ratios in different potential models.}
\label{tab2}
\begin{tabular}{lclclcl}
\hline\hline\\[-1.5 ex]
\multicolumn{1}{c}{Potential Model}                                                     &\multicolumn{1}{c} {$f_{B}$}     & \multicolumn{1}{c}{$f_{B^*}$} &\multicolumn{1}{c} {$f_{B_s}$}     & \multicolumn{1}{c}{$f_{B_s^*}$} & \multicolumn{1}{c}{$f_{B^*}/f_{B}$} & \multicolumn{1}{c}{$f_{B_{s^*}}/f_{B_s}$}
\\
\hline\\[-1.5 ex]
Martin \cite{martin}  &\multicolumn{1}{c} {206} &\multicolumn{1}{c} {227}                   &\multicolumn{1}{c}  {232} &\multicolumn{1}{c} {254} & \multicolumn{1}{c} {1.10} & \multicolumn{1}{c} {1.09} \\
Cornell \cite{cornell}                                                                                  &\multicolumn{1}{c} {193} &\multicolumn{1}{c} {211}                      &\multicolumn{1}{c} {219} &\multicolumn{1}{c} {239} & \multicolumn{1}{c} {1.09} & \multicolumn{1}{c} {1.09}          \\
Logarithmic \cite{log}                                                                              &\multicolumn{1}{c} {207} &\multicolumn{1}{c} {228}                     &\multicolumn{1}{c} {234} &\multicolumn{1}{c} {257} & \multicolumn{1}{c} {1.10} & \multicolumn{1}{c} {1.10}                  \\
Harmonic plus Yukawa \cite{harmonic}  
&\multicolumn{1}{c} {129} &\multicolumn{1}{c}    {138}                      &\multicolumn{1}{c} {174} &\multicolumn{1}{c}    {186} & \multicolumn{1}{c} {1.07} & \multicolumn{1}{c} {1.07}                       
\\
\hline\hline
\end{tabular}
\end{table}
We note that our predictions for the decay constants of $B$ mesons are more or less in agreement with the available experimental data ($f_B = 188(17)(18)$ MeV) \cite{patri}. The present predictions for the pseudoscalar and the vector mesons decay constants are important and  have many phenomenological implications especially in studying the $CP$ violation and in extracting the CKM matrix elements.




\end{document}